\documentstyle[prl,twocolumn,aps,epsfig]{revtex}
\begin{document}

\draft

\title{A Waveguide for Bose-Einstein Condensates}

\author{K. Bongs, S. Burger, S. Dettmer, D. Hellweg, J. Arlt, W. Ertmer, and K. Sengstock}
\address{Institut f\"ur Quantenoptik, Universit\"at Hannover, Welfengarten 1, 30167 Hannover, Germany}
\date{\today}
\maketitle

\begin{abstract}
We report on the creation of Bose-Einstein condensates of $^{87}$Rb in a specially
designed hybrid, dipole and magnetic trap. This trap naturally allows the coherent
transfer of matter waves into a pure dipole potential waveguide based on a doughnut beam.
Specifically, we present studies of the coherence of the ensemble in the hybrid trap and
during the evolution in the waveguide by means of an autocorrelation interferometer
scheme. By monitoring the expansion of the ensemble in the waveguide we observe a mean
field dominated acceleration on a much longer time scale than in the free 3D expansion.
Both the autocorrelation interference and the pure expansion measurements are in
excellent agreement with theoretical predictions of the ensemble dynamics.
\end{abstract}


The recent realization of Bose-Einstein condensation (BEC) in dilute atomic gases
\cite{BECs} has stimulated extensive studies on degenerate quantum gases. While most of
the experimental work so far has concentrated on 3D systems, there is growing interest in
systems with lower dimensionality leading to fundamentally different phenomena. The
specific properties of 1D quantum gases were recently studied theoretically and discussed
controversly, e.g., the existence of a pure condensate or a quasi condensate in a weakly
trapped 1D system \cite{Ketterle1D,Shlyapnikov}, the connection to the concept of a
Luttinger liquid \cite{Luttinger}, and the behaviour of density and phase fluctuations
\cite{Shlyapnikov}. The transfer of a Bose-Einstein condensate into a 1D system as
discussed here can thus provide important information, e.g., about the coherence
properties associated with the development of phase fluctuations \cite{Shlyapnikov}. A
wealth of new phenomena is also expected to occur in 1D collisional physics which may be
studied in the expansion of a dense ensemble transferred to a 1D or quasi 1D waveguide.
Waveguides with high transverse frequencies provide a tool for the experimental
realization of famous theoretical models such as a 1D gas of impenetrable bosons, the so
called Tonks gas \cite{Tonks,Olshanii}.

\begin{figure}[h]
   \begin{center}
   \parbox{8cm}{\epsfxsize 8cm\epsfbox{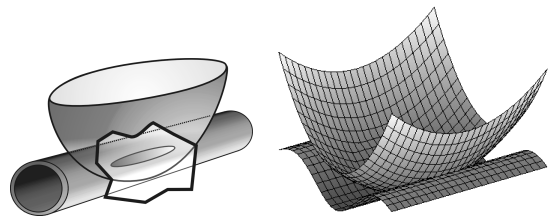} }
   \end{center}
   \caption{Outline of the DM-trap (a) Schematic view of the potential configuration.
   (b) Magnetic and dipole potential in the case of matched radial oscillation frequencies.}
   \label{hybridfalle}
\end{figure}

In this letter we report on the first experimental realization of the transfer of BECs
into a quasi 1D waveguide created by a blue detuned hollow laser beam. To transfer the
atoms coherently we have developed a new type of hybrid trap (Fig.\ref{hybridfalle}).
This combined dipole and magnetic (DM) trap consists of a waveguide added to the 3D
potential of a Ioffe type magnetic trap. It allows for a natural connection between the
magnetic trap and a pure 1D waveguide geometry created by a Laguerre-Gaussian
(TEM$_{01}^{*}$) laser beam. In our scheme, we directly obtain BEC by rf evaporation in
the DM-trap. By subsequently ramping down the magnetic trapping field of the DM-trap, the
BEC is transferred into the pure dipole potential waveguide and its dynamics is observed.
In particular the phase-coherence of the ensemble at the different stages of this process
is investigated with a Bragg interferometer scheme. As a result we find a phase-coherent
and mean field dominated acceleration in the waveguide on a time scale of $20\, $ms, much
longer than the mean field release time in the usual 3D expansion of a condensate.

As a class of waveguides, dipole potentials formed by blue detuned hollow laser beams are
well suited for gaining insight into different radial confinement regimes. By changing
the beam parameters they can be tuned from systems with a radial energy level spacing smaller
than the mean field energy of the ensemble to 1D Tonks gas systems with strong radial
confinement.

\begin{figure}[h]
   \begin{center}
   \parbox{7cm}{\epsfxsize 7cm\epsfbox{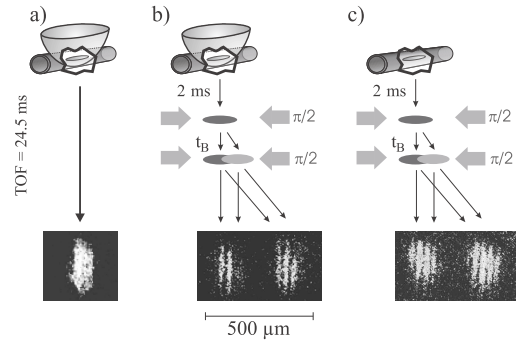} }
   \end{center}
   \caption{a) Absorption image of an atomic cloud after 24.5$\, $ms of time-of-flight
            for a BEC created in the DM-trap.
            b) Autocorrelation interferometer scheme and absorption image of the interference
            fringes for $t_B=3\, $ms and a total time-of-flight of 24.5$\, $ms for an ensemble
            released from the DM-trap.
            (c) Same as (b) for $t_B=1\, $ms and an ensemble stored in the waveguide for an
            evolution time of $6\, $ms.}
   \label{hybrid}
\end{figure}

Dipole potentials have previously been used to manipulate BECs. The creation of BECs in a
red detuned dipole and magnetic trap \cite{Zipfelfalle} and studies of spinor condensates
in a 3D red detuned dipole trap \cite{Ketterle-Spinor} as well as the transfer of a BEC
into an optical lattice \cite{Kasevich} were recently reported. Note however that a
focused red detuned laser beam creates a 3D trapping potential whereas the blue detuned
hollow TEM$_{01}^*$ laser mode discussed here represents an open 1D waveguide.
The radially symmetric intensity distribution,  $I(r)$, of a TEM$_{01}^*$ mode is given
by
\begin{equation}
I(r)=\frac{4 P r^{2}}{\pi r_0^{4}} \cdot e^{-2 \frac{r^2}{r_0^2}},
\end{equation}
where $P$ and $r_0$  are the laser power and the beam waist, respectively. With a power
of $P=1\, $W at 532$\, $nm and a beam waist of $r_0\approx 10\, \mu $m, a dipole
potential at the focal plane with a maximum value of $\sim 120\, \mu $K$\cdot k_B$ and a
transverse oscillation frequency of $\sim 6\, $kHz for $^{87}$Rb atoms can be realized.
Since the atoms are guided in the low intensity region of the light field, light
scattering which leads to decoherence is suppressed for atoms in the transverse ground
state by more than three orders of magnitude in comparison with the scattering rate in
the intensity maximum. For our measurements we adjusted the laser parameters such that
the mean radial oscillation frequency of 415$\pm 10\, $Hz matches the radial confinement
of the magnetic trap with an oscillation frequency of $400\pm10\, $Hz. The waveguide was
superimposed on the long axis of the magnetic potential (Fig.\ref{hybridfalle}) resulting
in a DM-trap with a total radial oscillation frequency of $ 576\pm14\, $Hz and an axial
oscillation frequency of 14$\, $Hz, solely due to the magnetic potential. Bose-Einstein
condensation in the DM-trap was obtained as follows: Approximately $10^9$ $^{87}$Rb atoms
were collected in a MOT. After their transfer to a cloverleaf magnetic trap, the atoms
were cooled  to a point just above the critical temperature ($T\approx1.5T_c$) by rf
induced evaporation. Then the light intensity of the waveguide potential was slowly
ramped up within $\approx 15\, $ms, transferring the ensemble into the DM-trap.
Subsequently a final rf evaporation ramp was applied, cooling the ensemble below $T_c$
and thus creating a Bose-Einstein condensate of up to $2\times10^5$ $^{87}$Rb atoms in
the DM-trap. Note that the DM-trap still allows rf evaporation of atoms via the waveguide
axis. We observe a lifetime of the BEC in the DM-trap on the order of one second which is
comparable to the BEC lifetime observed in our pure magnetic trap. To confirm the onset
of BEC we observe the sudden increase of optical density of the sample and its
anisotropic expansion in a time-of-flight measurement as shown in Fig.\ref{hybrid}(a).

To investigate the phase-coherence of the ensemble in the DM-trap we use a simple two
beamsplitter interferometer scheme. We also use this coherence detector to study the
phase-coherence of the ensemble after transfer into the waveguide as discussed later.

The interferometer consists of only two $\frac{\pi }{2}$ Bragg pulses separated by a time
$t_B$, applied after switching off the trapping potentials as indicated in
Fig.\ref{hybrid}(b). The beamsplitters are realized by $\frac{\pi }{2}$ Bragg pulses
\cite{Ketterle-Bragg,Phillips-Bragg} of counterpropagating laser beams aligned parallel
to the axis of the waveguide, which split an incoming wavepacket into a coherent
superposition of two wavepackets with velocities differing by $2\hbar k/m=11.7\, $mm/s.
The two laser frequencies are detuned by $\approx 6\, $GHz from the $^{87}$Rb-line at
780$\, $nm with a fixed frequency difference set to match the Bragg condition. The two
wavepackets created by the first $\frac{\pi }{2}$-pulse separate in space during the free
evolution time $t_B$ by an amount $\Delta x = \frac{2\hbar k}{m}t_B$. The second
$\frac{\pi }{2}$-pulse then recombines the partially overlapping clouds in both exit
ports of the interferometer, which again differ in momentum by $2\hbar k$. The
displacement leads to interference fringes in both exit ports due to the additional phase
distribution caused by the expansion of the condensate, similar to the autocorrelation
measurements in the $\frac{\pi }{2}-\pi-\frac{\pi }{2}$ geometry of
\cite{Phillips-Autokorrelation}. 
The observed spacing of the interference fringes can be understood as follows: The
condensate expands mainly due to the conversion of mean field energy into kinetic energy.
Since the mean field energy, $U_{mf}$, is proportional to the parabolic density
distribution, the local acceleration $\dot v \propto \nabla U_{mf}$ depends linearly on
position. As a consequence the ensemble keeps a parabolic density distribution during
expansion while acquiring a velocity field which linearly increases with position, $v(x)
= \alpha (t)x$. The velocity gradient is given by $\alpha(t)=\dot \lambda (t)/\lambda
(t)$ with the scaling parameter $\lambda (t)$ defined in \cite{CastinDum} indicating the
size of the condensate relative to its size in the trap. The local velocity difference
between the two overlapping clouds thus does not depend on position but only on the
displacement and on time $\Delta v = \alpha (t) \Delta x$. The observed interference
pattern therefore corresponds to the interference of two plane matter waves with a
relative velocity $\Delta v$. The corresponding fringe spacing $d = \frac{h}{m\Delta v }$
can be used to deduce the velocity gradient of the ensemble
\begin{equation}\label{expansion_parameter}
  \alpha (t) = \frac{h}{md\Delta x} \, .
  \label{dieda}
\end{equation}
Note that the above discussion assumes that all mean field energy is converted to kinetic
energy before the interferometer sequence. In our experiment the time-of-flight prior to
the first interferometer pulse is chosen to be 2$\, $ms in which more than $95\% $ of the
mean field energy is converted into kinetic energy.

We now turn to our measurements of phase-coherence for 3D ensembles in the DM-trap and
during expansion in the quasi 1D waveguide. We obtained the interference pattern of the
BEC produced in the DM-trap by taking absorption images of the density distribution after
the interferometer sequence and a total time of flight of $24.5\, $ms
(Fig.\ref{hybrid}(b)). The time between the pulses was varied from $t_B=1$ to 4ms
resulting in a displacement of the interfering clouds of $\Delta x = 12$ to $48\, \mu$m.
The largest displacements roughly correspond to half the condensate size of $\approx
100\, \mu $m. For all these displacements high contrast, equally spaced interference
fringes were observed, demonstrating both, the coherence of the ensemble on these length
scales and the validity of the above model for the 3D BECs in the DM-trap.

In a further set of experiments, the transfer of the Bose-Einstein condensed ensemble to
the waveguide and the subsequent evolution of the atomic ensemble inside the waveguide
were investigated.
\begin{figure}[h]
   \begin{center}
   \parbox{8cm}{ \epsfxsize 8cm \epsfbox{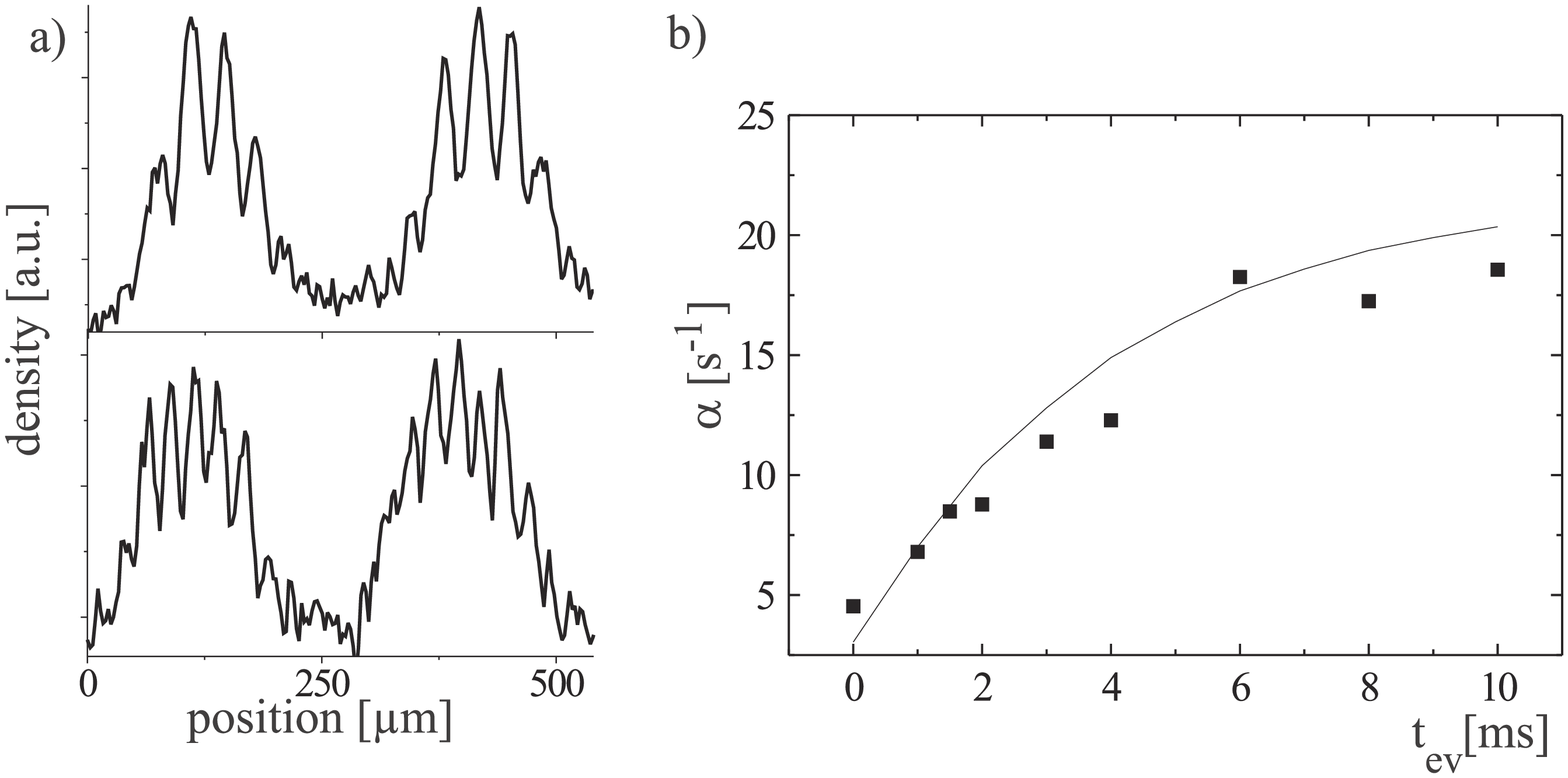} }
   \end{center}
   \caption{(a) Interference signals from guided ensembles for evolution times of
            $4\, $ms (top) and $6\, $ms (bottom) in the waveguide. (b) Velocity gradient
            versus guiding time deduced from measurements of the fringe spacing (dots)
            in comparison to the numerical calculation (solid line).}
   \label{coherent-donut}
\end{figure}
Again, in a series of measurements the $\frac{\pi}{2}-\frac{\pi}{2}$ pulse interferometer
scheme was now applied to an ensemble transferred from the DM-trap into the waveguide for
different evolution times $t_{ev}$ in the waveguide. For this purpose the natural
connection of the DM-trap to the waveguide was demonstrated by adiabatically lowering the
magnetic field, thus releasing the ensemble into the waveguide. Figure~\ref{hybrid}(c)
shows an example of the resulting interference fringes. Figure~\ref{coherent-donut}(a)
shows examples of cross sections through the interference signals. For these measurements
the time between the Bragg pulses was set to 1ms, resulting in a displacement of the
wavepackets of $\approx12\, \mu$m. For the first time, these interference signals clearly
document the axial phase-coherence of an ensemble after the transfer of a BEC into a
quasi 1D waveguide and an additional evolution time. As another important result, the
equal spacing of the interference fringes confirms the linear velocity distribution
predicted by the model given above (Eqn.\ref{dieda}) for the 1D expansion in the
waveguide.

The decrease of fringe spacing for increased evolution times in the waveguide
demonstrates an increasing velocity spread of the wavepacket as mean field energy is
converted into kinetic energy within the waveguide. We observe a decrease in fringe
spacing with time even for evolution times of $10\, $ms in the waveguide, i.e., the atoms
are still accelerated by the mean field energy. These measurements clearly show that the
conversion of mean field energy into kinetic energy is more than one order of magnitude
slower than in 3D due to the reduced dimensionality inside the waveguide.

From the regular fringe spacing we deduced the spatial velocity gradient of the ensemble
for different evolution times. Fig.\ref{coherent-donut}(b) compares this data to the
result of a numerical calculation of the parameter $\alpha(t)$ according to
\cite{CastinDum}. Note that the calculation does not include any free parameters, nor
depends on the particle number. The good agreement clearly demonstrates the consistency
of the ensemble dynamics with a mean field dominated and phase coherent expansion inside
the waveguide and the applicability of the scaling laws to the quasi 1D regime.
\begin{figure}[h]
   \begin{center}
   \parbox{5cm}{\epsfxsize 5cm\epsfbox{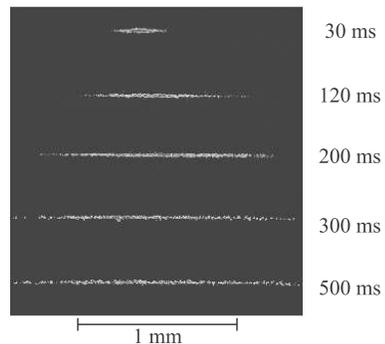} }
   \end{center}
   \caption{Atoms from a BEC loaded to a doughnut waveguide with different
            evolution times inside the waveguide.}
   \label{tof-lang}
\end{figure}
The above interferometer scheme is well suited to study the axial expansion of the
ensemble inside the waveguide within the first ten milliseconds. For longer guiding
times, when the ensemble density is significantly reduced we have directly studied the
expansion by absorption imaging without an additional time-of-flight.
Figure~\ref{tof-lang} shows examples for these measurements for evolution times up to
$500\, $ms. On time scales above $40\, $ms the conversion of mean field energy into
kinetic energy is nearly complete and the ensemble is expected to expand with constant
velocity keeping its parabolic density distribution. We indeed found the predicted
parabolic density distribution confirmed for evolution times up to $150\, $ms from cross
sections through images as shown in Fig.\ref{tof-lang}. These measurements are limited
only by our field of view as no reliable fits to the data can be obtained as soon as the
imaged length of the ensemble corresponds to the size of our CCD array.

An estimate of the maximum velocity at the edge of the cloud after an expansion in the
waveguide can be obtained as follows: After long expansion times in the waveguide the
ensemble width is much larger than its initial width. In this limit the maximum velocity
at the edge of the cloud can be calculated from simple kinematics and is given by
\begin{equation}
   v_{max} = \sqrt{\frac{10 E_{\rm kin}}{m}}
\end{equation}
where $E_{kin}$ is the kinetic energy per particle. If all internal energy is converted
in kinetic energy, $E_{kin}$ can be replaced by the initial internal energy $E_{int}$.
The internal energy per particle in the Thomas-Fermi limit is well known \cite{Baym} and
is given by $E_{\rm int} = 0.46 k_B N^{2/5}\, $nK for our experimental parameters. For
N=$5\times10^4$ atoms we find $v_{max}= 5.8 \, $mm/s in good agreement with the velocity
deduced from our experimental data $v_{max}= 5.9 \, $mm/s. This again confirms the mean
field dominated expansion of the cloud in the waveguide.
\begin{figure}[h]
   \begin{center}
   \parbox{7cm}{
   \epsfxsize 7cm
   \epsfbox{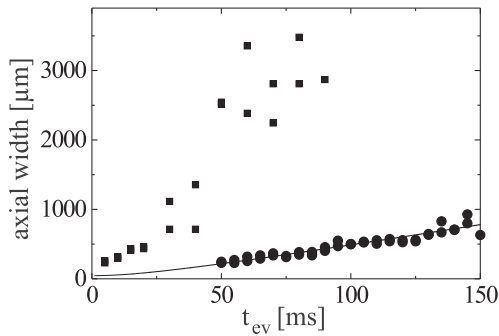}}
   \end{center}
   \caption{Longitudinal expansion of BECs loaded into the waveguide.
            Squares show the Gaussian half-width of
            ensembles transferred from the magnetic trap. Circles represent the Thomas-Fermi
            half-width of ensembles adiabatically
            transferred from the DM-trap and the solid line corresponds to a theoretical prediction for adiabatic
            loading conditions with $N=5\times10^4$ atoms.}
   \label{broadening}
\end{figure}
Finally, we have compared different loading mechanisms of the waveguide.
Figure~\ref{broadening} shows a comparison of the expansion of atomic clouds loaded from
the DM-trap to the expansion of clouds loaded directly from a BEC created in the
cloverleaf magnetic trap. The latter was done by instantly switching on the doughnut beam
and switching off the magnetic trap in $200\mu$s. For direct loading from the magnetic
trap, the system expands to several millimeters in length within $\approx 100$\,ms, which
indicates the non-adiabaticity of the direct transfer process. In contrast, the cloud's
expansion for loading from a BEC first created in the DM-trap shows excellent agreement
with the theoretical prediction~\cite{CastinDum} of mean field dominated expansion (solid
line in Fig.~\ref{broadening}). In this case the atomic density distribution spreads out
over an axial region of $\approx 1$\,mm within $150\,$ms, and the cigar-shaped BEC
transforms to a straight hair of quantum gas.

In conclusion we have presented three important results, paving the way to study 1D
quantum gases. First, we have observed Bose-Einstein condensation of an ensemble of
$^{87}$Rb atoms in a hybrid trap realized by the combined potential of a Ioffe type
magnetic trap and a blue detuned dipole waveguide potential. Second, the coherence of the
ensemble in the trap and in the waveguide was measured with a simple autocorrelation
interferometer scheme. Third, we investigated the transfer process of Bose-Einstein
condensates into a blue detuned dipole waveguide and studied the subsequent evolution of
the ensemble in a quasi 1D waveguide. In these experiments we demonstrated that a fully
coherent transfer is possible by our scheme and observed a mean field dominated expansion
of the ensemble for adiabatic loading conditions.

These investigations open a path for future studies of the different regimes of 1D
quantum gases. The realization of a Tonks gas of impenetrable Bosons seems feasible in
our geometry. After loading the dipole waveguide potential can be increased adiabatically
to reach the 1D regime in which transverse degrees of freedom are fully frozen out
($\omega_r\approx10\, $kHz). As additional magnetic fields can be superimposed on the
dipole potential, the scattering length $a$ can be tuned using Feshbach resonances to
achieve a suitable value for the Tonks gas regime. Also, e.g. spinor condensates
\cite{Ketterle-Spinor} and dark solitons \cite{Phillips-Solitons,Solitons} can be
investigated in 1D waveguide geometries. The waveguide can easily be closed by additional
light field mirrors \cite{bouncingBEC} or a superimposed weak longitudinal trapping
potential. This opens opportunities for a first realization of a guided matter wave
interferometer.

This work is supported by SFB\,407 of the {\it Deutsche Forschungsgemeinschaft}.


\begin{references}

\bibitem{BECs} M. J. Anderson, J. R. Ensher, M. R. Matthews, C. E. Wieman, and E.
A. Cornell, Science {\bf 269}, 198 (1995); K. B. Davis, M.-O. Mewes, M. R. Andrews, N. J.
van Druten,  D. S. Durfee, D. M. Kurn, and W. Ketterle, Phys. Rev. Lett {\bf 75}, 3969
(1995);  C. C. Bradley, C. A. Sackett, and R. G. Hulet, Phys. Rev. Lett. {\bf 78}, 985
(1997); D. Fried et al., Phys. Rev. Lett. {\bf 81}, 3811 (1998)

\bibitem{Ketterle1D} N. van Druten and W. Ketterle, Phys. Rev. Lett. {\bf 79}, 549 (1997)

\bibitem{Shlyapnikov} D. Petrov, G. V. Shlyapnikov and J. Walraven, cond-mat/0006339 and
references therein

\bibitem{Luttinger} H. Monien, M. Linn and N. Elstner, Phys. Rev. A {\bf 58}, R3395 (1998)

\bibitem{Tonks} L. Tonks, Phys. Rev. 50, 955 (1936)

\bibitem{Olshanii} M. Olshanii, Phys. Rev. Lett. {\bf 81}, 938 (1998)

\bibitem{Zipfelfalle} D. M. Stamper-Kurn et al., Phys. Rev. Lett. {\bf 80}, 2027 (1998)

\bibitem{Ketterle-Spinor} J. Stenger et al., Nature {\bf 396}, 345 (1998)

\bibitem{Kasevich} B. Anderson and M. Kasevich, Science {\bf 282}, 1686 (1998)

\bibitem{Ketterle-Bragg} J. Stenger et al., Phys. Rev. Lett. {\bf 82}, 4569 (1999)

\bibitem{Phillips-Bragg} E.W. Hagley et al., Phys. Rev. Lett. {\bf 83}, 3112 (1999)

\bibitem{CastinDum} Y. Castin and R. Dum, Phys. Rev. Lett. {\bf 77}, 5315 (1996)

\bibitem{Phillips-Autokorrelation} J. Simsarian et al, cond-mat/0005303

\bibitem{Baym} G. Baym and C. Pethick, Phys. Rev. Lett. {\bf 76}, 6 (1996)

\bibitem{Phillips-Solitons} J. Denschlag et al., Science {\bf 287}, 97 (2000)

\bibitem{Solitons} S. Burger et al., Phys. Rev. Lett. {\bf 83}, 5198 (1999)

\bibitem{bouncingBEC} K. Bongs et al., Phys. Rev. Lett. {\bf 83}, 3577 (1999)

\end{references}
\end{document}